# Plasmonically Enhanced Flexural-Mode AlScN Nanoplate Resonator as Uncooled and Ultrafast IR Detector with High Responsivity

Aurelio Venditti, Walter Gubinelli, Enise F. Altin, Luca Colombo, Pietro Simeoni, Benyamin Davaji, and Matteo Rinaldi[*]

*Institute for NanoSystems Innovation, Northeastern University, Boston MA (USA)*

E-mail: vendetti.a@northeastern.edu

Phone: +1 (857) 2896101

## Abstract

This letter introduces a novel class of miniaturized, uncooled, and ultra-fast infrared (IR) resonant thermal detectors (RTDs) based on 30%-doped Aluminum Scandium Nitride (AlScN) nanoplates. Exploiting high electromechanical coupling, good thermal properties, and enhanced and selective IR absorption, the presented device aims to demonstrate significant advancements over the state-of-the-art IR RTDs. This single pixel combines compact footprint, high spectral selectivity and responsivity, reduced noise, and fast thermal response, allowing for the potential development of innovative IR thermal imagers through multi-pixel integration. The flexural nature of the actuated resonance mode eventually enables an interferometric optical readout, paving the way towards achieving extremely low Noise Equivalent Power levels.

These results demonstrate a high IR responsivity of around 130 ppt/pW, a thermal time constant of around 330 µs, and a large out-of-plane displacement. This work represents the first experimental integration on a resonating platform of plasmonic absorbers that utilize AlScN as dielectric layer.

# Introduction

Conventional uncooled infrared (IR) imagers rely almost exclusively on VOx or a-Si-based microbolometer arrays, which are fabricated with complementary metal oxide semiconductor (CMOS)-compatible processes. However, these devices suffer from inherent trade-offs between Noise Equivalent Power [1,2] (NEP), commonly in the 100s of pW/Hz$^{1/2}$, and thermal time constant [3] ($\tau$), which is typically around 10s of ms. Efforts to improve one metric frequently result in the degradation of the other because of their opposite dependence on the thermal conductance [4], limiting overall sensor performance. Single-pixel pyroelectric passive IR (PIR) detectors and thermopiles simplify the read-out electronics but are about an order of magnitude less sensitive (NEP usually around 1s nW/Hz$^{1/2}$), with comparable thermal time constants, but larger footprints still in the millimeter scale [5,6]. Additionally, all these thermoresistive, pyroelectric, or thermoelectric-based approaches typically lack spectral selectivity due to their intrinsic broadband spectral response. Among the various emerging technologies for uncooled thermal detectors (TDs), nanoplate resonators (NPRs) based on Aluminum Nitride (AlN) have demonstrated great potential, offering notable improvements in terms of detection limit and response time, together with extremely low power consumption. Such performance enhancement has been obtained by scaling both the thickness and lateral dimensions of the piezoelectric film, while simultaneously preserving high values of all the other figures of merit, including resonance quality factor, transduction efficiency, and thermal resistance. Recent demonstrations in contour-mode resonators with full electrical drive and readout have achieved a NEP ≈ 86 pW/Hz$^{1/2}$ and a $\tau$ ≈ 166 $\mu$s [7]. To overcome the NEP limitations imposed by the noise originating from the electronic readout in prior implementations, this work exploits out-of-plane flexural vibration modes in an NPRTD, aiming for its integration with a shot-noise-limited interferometric optical readout to potentially achieve ultra-low NEP in the order of single-digit pW/Hz$^{1/2}$. Furthermore, an aggressive miniaturization of the lateral size of the plate enables extremely small $\tau$. Finally, this study demonstrates for the first time the feasibility of integrating plasmonic metamaterial absorbers on top of the NPR using AlScN as dielectric layer, resulting in the significant enhancement of the TD responsivity and selectivity to IR radiation [8].

Table 1: Uncooled Thermal IR Detectors Overview.

| Detector Type | Active Area [$\mu m^2$] | Responsivity | Thermal Time Constant [$ms$] | Noise Equivalent Power [pW/Hz$^{1/2}$] |
|---|---|---|---|---|
| Microbolometer[9] | 17 x 17 | 10,000s V/W | 10s | 100s |
| PIR Detector[5] | 1000 x 1000 | 1,000s V/W | 10s | 1,000s |
| Thermopile[6] | 600 x 600 | 100s V/W | 10s | 1,000s |
| Contour Mode Resonator[7] | 20 x 22 | 70 ppt/pW | 0.166 | 86 |
| Flexural Mode Resonator[*] | 23 x 13 | 132 ppt/pW | 0.329[a] | 1s[b] |

[*] This work. [a] 3D FEM simulations. [b] Optical readout.

## Device Design and Simulations

The architecture of the proposed NPRTD leverages a suspended Sc-doped AlN (AlScN) nanoplate structure shown in Figure 1a, designed to efficiently enable a strong out-of-plane displacement at resonance to eventually implement an interferometric optical readout. This perspective led to a nanoplate design consisting of a double 100 nm-thick 30%-doped AlScN layer sandwiched between 20 nm-thick Platinum (Pt) electrodes and connected to the substrate through four anchors. The geometry, location, and routing of the electrodes are selected to efficiently excite symmetric flexural modes, as illustrated in the inset of Figure 1b. In particular, the out-of-plane deformation is induced by an in-plane compressive and tensile stress in the bimorph AlScN plate due to its $e_{31}$ piezoelectric coefficient, combined with an in-plane clamping originating from the anchoring system [10,11]. The degradation of electromechanical properties associated with reduced crystallinity of the piezoelectric films and increased parasitic losses, which is typically caused by the scaling of the device dimensions, is mitigated by an enhanced piezoelectric coefficient provided by high-Sc-doping as well as by the proper placement of both electrodes and anchors [12]. Sc-doping and high metallization are also leveraged to improve the sensing capabilities by increasing the responsivity to IR radiation. It has been demonstrated that increasing Sc-doping heavily decreases the thermal conductivity [13], thereby boosting the thermal resistance of the material. Moreover, the temperature coefficient of frequency increases with increasing Sc-doping [14], and the high ratio between the Pt and AlScN layers thickness further boosts the thermal responsivity of the device.

A ButterworthVan Dyke (BVD) model was employed to fit the admittance response shown in Figure 1b, which was obtained through a frequency-domain 3D Finite Element Modeling (FEM) simulation carried out incorporating a model for AlScN with Sc-doping level-dependent piezoelectric coefficients, mass density, and relative permittivity [15]. The key electromechanical parameters of the device, including static capacitance ($C_0$), resonance frequency ($f_s$), and electromechanical coupling coefficient ($k_t^2$), were extracted. In particular, the targeted mode exhibits a resonance frequency $f_s$ near 16.6 MHz, with an excellent $k_t^2$ of approximately 1.9% and a low $C_0$ of 0.26 pF. Additionally, the thermal properties under vacuum conditions were assessed via time-domain 3D FEM simulations under uniform IR irradiation. By fitting the temperature rise in the AlScN layers over time induced by a constant and uniform IR power impinging on the surface of the device (Figure 1c), a high thermal resistance ($R_{th}$) of 1.2 K/µW and a small thermal capacitance ($C_{th}$) of 277 pJ/K were estimated, leading to a short $\tau$ of the nanoplate equal to approximately 330 µs. Its reduction can be obtained by further shrinking both the lateral dimensions of the nanoplate and thinning the thickness of the AlScN layers.

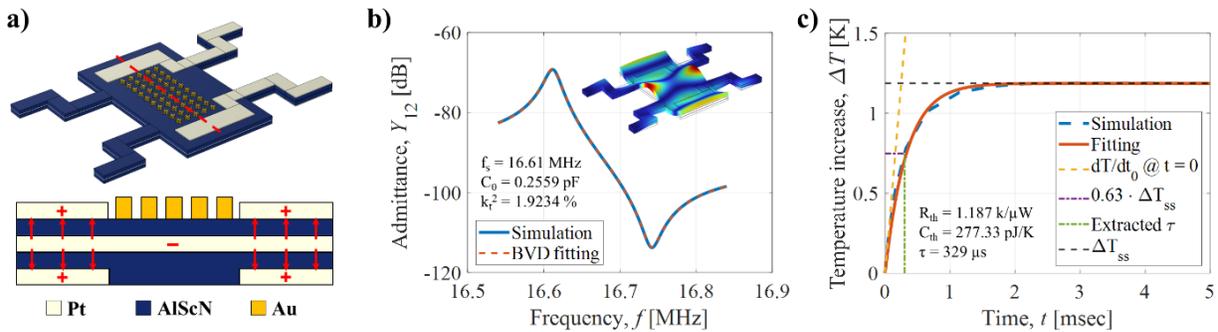

Figure 1: a) 3D FEM model of the device and its corresponding cross-sectional view illustrating the material stack and the electrode configuration; b) Simulated admittance response obtained via 3D FEM and corresponding BVD model fitting of the resonance with its mode shape achieving large out-of-plane displacement; and c) Simulated thermal transient response obtained via 3D FEM simulation and corresponding exponential fitting to evaluate the thermal properties.

Finally, in order to simultaneously improve the responsivity to IR detection of the sensor and introduce spectral selectivity, a plasmonic metamaterial absorber was properly designed, exploiting the same material stack of the nanoplate to integrate a Metal-Insulator-Metal (MIM) structure, while maintaining the compatibility with the underlying AlScN NPRTD without compromising its electromechanical performance. This metasurface consists of a 2D periodic array of 100 nm-thick Gold (Au) cross-shaped nanostructures, whose unit cell is carefully dimensioned and spaced to achieve high and spectrally selective IR absorptance in a targeted wavelength range (Figure 2a) [16]. While Silicon Dioxide (*SiO₂*) is conventionally employed as dielectric layer in MIM structures, its positive Temperature Coefficient of Frequency (TCF) negatively impacts the overall TCF of the detector. For this reason, alternative dielectric materials such as AlN and AlScN have been explored [17,18]. In particular, in this work AlScN was selected and integrated on such a small resonating platform for the first time to further improve the overall thermal properties of the detector with respect to previous demonstrations. The absorptance spectrum ($\eta$) of this structure was assessed through a 3D FEM simulation and is shown in Figure 2b, where a central wavelength of 5.3 μm was targeted for testing purposes. Additionally, the electric and magnetic field cross-sectional distributions in the material stack at the absorptance peak are shown in Figure 2c and Figure 2d, respectively. In particular, the electric field is highly enhanced at the metal-dielectric interface, while the magnetic field is strongly confined in the dielectric layer when the localized plasmon polariton is generated at the plasmonic resonance. Therefore, this metasurface is engineered to efficiently convert the electromagnetic energy of the impinging IR radiation into heat only in a targeted spectral range, inducing an increase in temperature inside the nanoplate that is larger than the case where the intrinsic absorption of the materials is merely exploited.

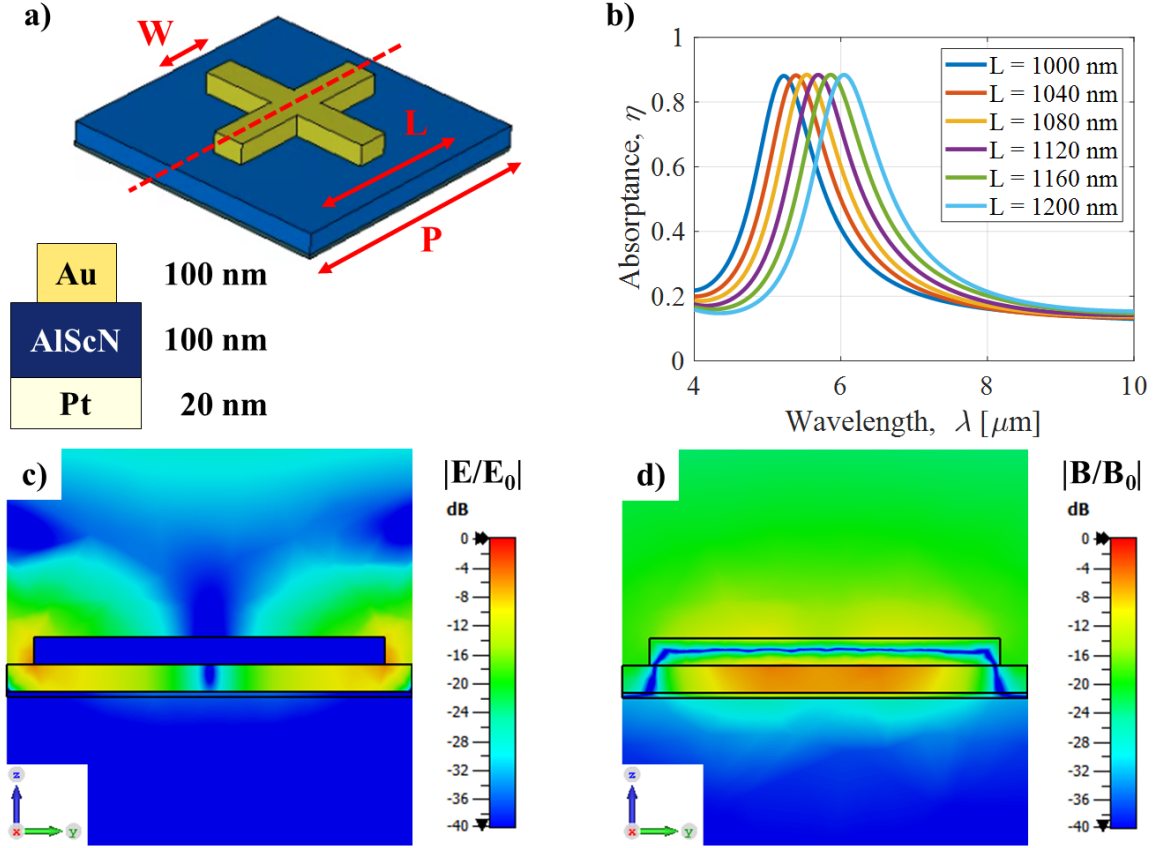

Figure 2: a) 3D FEM model and cross-sectional view of the plasmonic absorber unit cell; b) Simulated absorptance spectra obtained via 3D FEM and its tuning obtained by sweeping the length L of the cross, while keeping both its periodicity P and width W fixed to 1.5 μm and 300 nm, respectively; c) 3D FEM simulated electric field cross-sectional distribution at resonance, showing enhancement at the metal-dielectric interface; and d) 3D FEM simulated magnetic field cross-sectional distribution at resonance, highlighting strong field confinement within the dielectric layer.

## Experimental Results and Discussion

The device was fabricated through a six-mask process, schematically depicted in Figure 3a. Standard microfabrication processes were used, and an Evatec CLUSTERLINE® 200 deposition tool was used to deposit 30%-doped AlScN piezoelectric thin films. A tilted-view Scanning Electron Microscope (SEM) image of the fabricated device illustrating its lateral dimensions is shown in Figure 3b, while Figure 3c highlights the patterned top metasurface.

A 2-port measurement under vacuum allowed the extraction of the admittance response shown in Figure 4a, which was performed after standard short-open-load calibration of the Keysight N5221A Vector Network Analyzer (VNA) at room temperature and atmospheric pressure. The results demonstrate a good agreement with the 3D FEM simulated response, with the parameters assessing the electromechanical performance of the device extracted by modified BVD (mBVD) fitting. The fitted values of $f_s$, $C_0$, quality factor (Q), and $k_t^2$ are shown in the inset of Figure 4a, proving the effectiveness of the proposed design. A TCF of approximately -70 ppm/K was extracted from repeated admittance measurements conducted across a wide temperature range controlled using a heated stage (Figure 4b). This relatively high TCF, which exceeds values reported in previous demonstrations [17], is attributed to the combined effects of high-density metallization and the introduction of Sc-doping in the AlN layer used as the dielectric material within the MIM structure. However, higher values of TCF were previously achieved on the same material stack with a similar device [18]. The reason a different value was observed on this device is likely the different mode shape of the excited resonance and its distribution on the material stack.

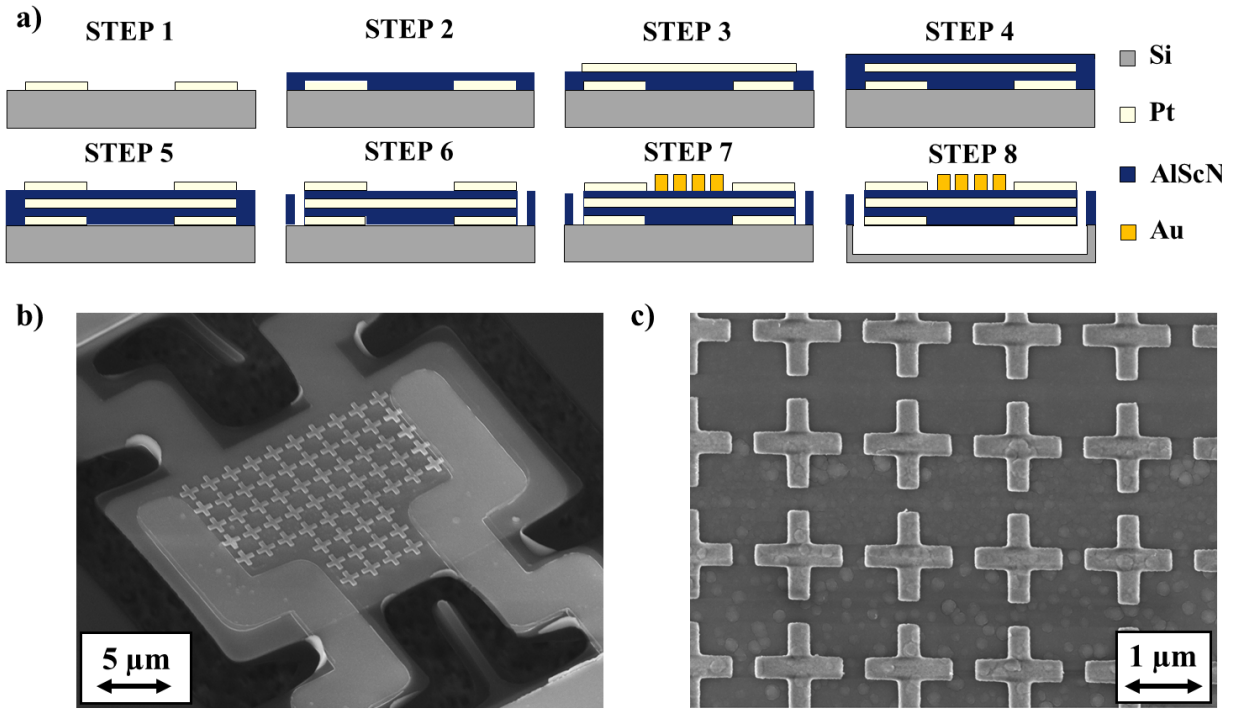

Figure 3: a) Overview of the main steps of the microfabrication process: 1) Sputtering and patterning by lift-off of Pt bottom electrode; 2) Reactive sputtering of the first AlScN layer; 3) Sputtering and patterning by lift-off of Pt intermediate electrode; 4) Reactive sputtering of the second AlScN; 5) Sputtering and patterning by lift-off of Pt top electrode; 6) Dry etching to define release pits and anchors; 7) Wet etching to open vias followed by evaporation and patterning by lift-off of Au absorbers and pads; 8) Device releasing through *XeF*$_2$ dry etching (or alternatively through deep reactive ion etching for optical interferometric readout compatibility); b) Tilted SEM image of the fabricated device; and c) SEM picture of the patterned plasmonic absorber array.

An IR laser with a central emission wavelength equal to 5.3 µm was then included in the optical setup shown in Figure 4c to experimentally assess the IR sensing performance of the device under test (DUT). To achieve this, the beams of both the IR laser and a red laser pointer were first aligned along the same straight line, then combined through a reflective prism to visually keep track of the IR beam over the whole optical setup, and finally pointed at the DUT. Moreover, a mechanical chopper was inserted in order to automatically perform an ON/OFF modulation of the IR radiation impinging on the DUT at a stable and known frequency. The variations in the admittance amplitude at the frequency of maximum admittance slope of the device between its resonance and antiresonance were monitored using the VNA in continuous wave (CW) mode. A frequency shift $\Delta f$ of the resonance of the DUT equal to 26.8 kHz was extracted by dividing the measured change in admittance amplitude by its slope (see Figure 4d). Prior to the measurements, the optical power density emitted by the IR laser was characterized using a mercury-cadmium-telluride photodiode power sensor (Thorlabs S180C) and applying the knife-edge technique to estimate the beam radius, which was then confirmed by varying the aperture of an iris diaphragm placed in front of the power meter. Estimating an IR optical power impinging on the device $P_{IR}$ of around 12.16 µW, an IR responsivity of 2.2 Hz/nW was calculated by dividing the induced frequency shift by the delivered IR optical power:

$$R_s = \frac{f_s}{P_{IR}} \approx f_s \cdot \eta \cdot |TCF| \cdot R_{th} \tag{1}$$

By normalizing the extracted responsivity with respect to the central resonance frequency of the device, a value of around 132 ppt/pW was extracted, which is almost two times higher than previous demonstrations with devices having similar size [7], where laterally vibrating resonators (LVRs) with resonating frequencies almost two orders of magnitude higher were used, and comparable to other LVRs demonstrated by Gülseren with a much bigger footprint [19].

Among all the different noise sources, the fundamental limit to frequency stability of an NPRTD is given by the thermomechanical noise originating from the thermally driven random motion of the nanoplate. The thermomechanically-induced frequency noise spectral density $f_n$ can be calculated as follows [20]:

$$f_n = \sqrt{\frac{k_B T}{4 P_c} \frac{f_s}{Q}} \qquad (2)$$

where $k_B$ denotes the Boltzmann constant, $T$ is the operating temperature (here assumed to be the ambient one), and $P_c$ represents the Radio Frequency (RF) driving power level applied to the device with the VNA (here set to -35 dB, as it is the maximum RF power level handled by the resonator without showing a non-linear behavior). According to the presented experimental results, an NEP associated with the thermomechanical noise of around 0.57 pW/Hz$^{1/2}$ is calculated as the ratio between the theoretically-calculated thermomechanically-induced noise fluctuations and the experimentally extracted responsivity:

$$NEP = \frac{f_n}{R_s} \qquad (3)$$

This value represents the ultimate limit eventually achievable with a quantum-limited integrated optical readout.

A Digital Holographic Microscope (DHM) was employed to experimentally verify the symmetry of the mode shape associated with the fitted resonance. This was performed by measuring the displacement at two points of the nanoplate symmetric with respect to its center and close to the theoretical points of maximum displacement according to the 3D FEM simulations, as shown in Figure 4e. Under a 300 mV peak-to-peak voltage excitation, an out-of-plane peak-to-peak displacement of approximately 70 nm was recorded for both the probed points without any phase offset (Figure 4f), confirming the induced symmetric motion of the nanoplate. Moreover, almost zero displacement was recorded at the center of the plate, thereby confirming the mode shape previously simulated by the 3D FEM model.

The plasmonic absorbers were both fabricated on the nanoplate as well as on larger test structures having the same material stack. A Fourier Transform Infrared Spectroscopy (FTIR) microscope (Bruker LUMOS) was used in reflection mode to measure their IR absorption efficiency, similar to what was previously done [18]. While the absorptance of the test structures (Figure 5a) matched well the simulated results shown in Figure 2b, FTIR measurements taken directly on the devices turned out to be challenging because of the small size of the nanoplate due to the limitations of the instrument. Nevertheless, Figure 5b shows how, despite the presence of the Pt electrodes and the non-infinite array of the metasurface, which was limited to the repetition of a few unit cells by the plate size, the spectral selectivity to IR is still conserved, while a huge degradation of the absorption peak height has been recorded. This can be attributed to the plate-limited size of the plasmonic absorbers array, as well as to the bending of the nanoplate after its releasing. Considering an absorption efficiency of 38% at the emission wavelength of the IR laser used in the optical setup built to characterize the IR radiation detection capabilities of the device, a thermal responsivity of approximately 347 ppt/pW is evaluated.

## Conclusions

The experimental results reported in this letter showcase the promise of a new class of uncooled thermal IR detectors featuring high sensitivity, low noise, and extremely fast response time. A 30%-doped AlScN nanoplate exhibiting extremely large out-of-plane displacement at resonance was demonstrated and integrated with IR plasmonic absorbers to be functionalized as an extremely selective and sensitive IR detector. Future developments include backside release of the devices to track the resonators frequency shift by optical interferometry.

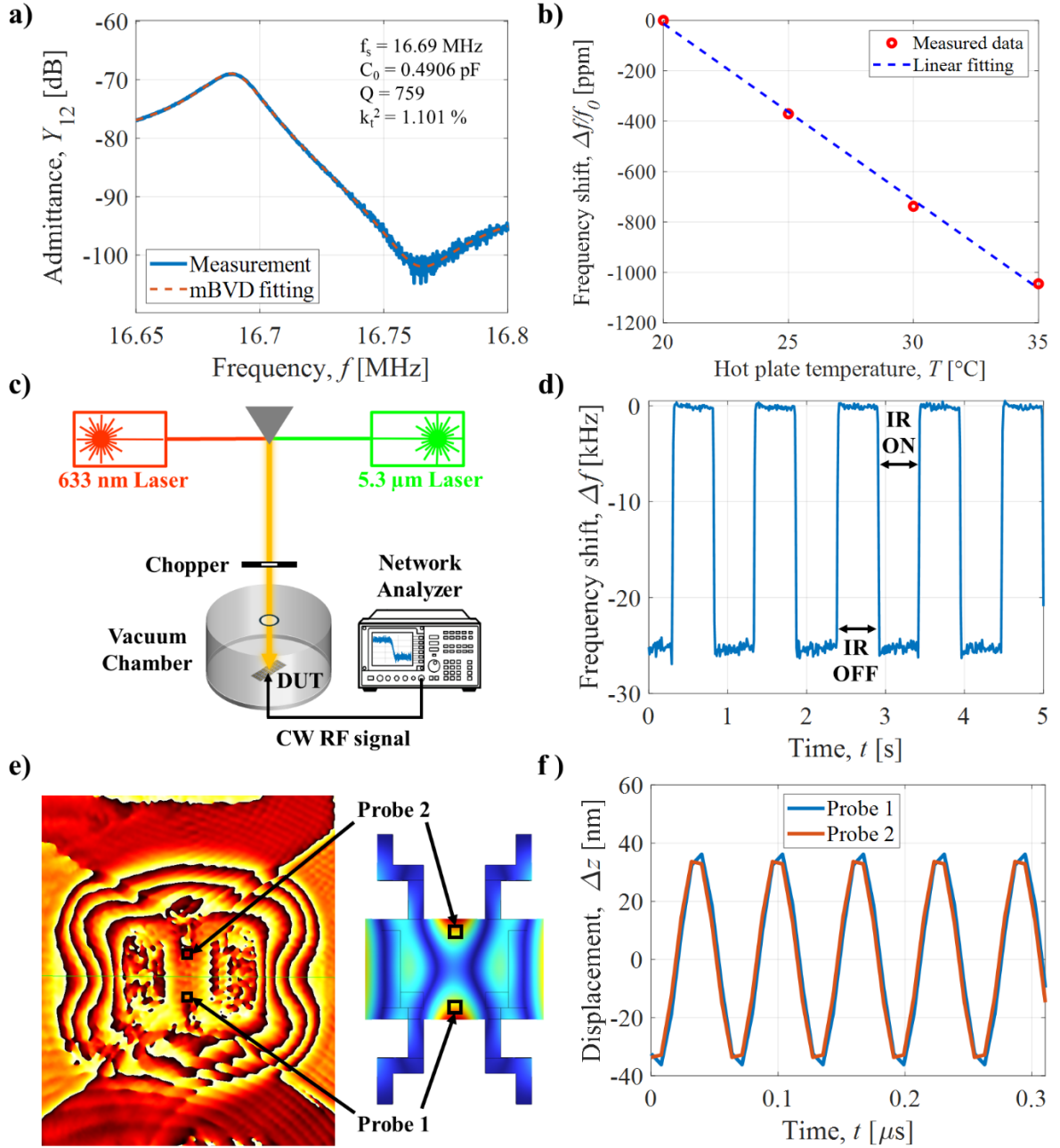

Figure 4: Experimental characterization of the device electromechanical response and its capability to detect IR radiation: a) Measured admittance response under vacuum conditions using a VNA; b) Extraction of the TCF from the slope of the normalized resonance frequency shift as a function of the temperature; c) Schematic of the experimental setup built to monitor the response of the device to IR radiation; d) Shift of the resonance of the device recorded at the frequency of maximum slope of the admittance curve recorded during the ON-OFF modulation of the IR source; e) Comparison between the mode shape obtained by the 3D FEM simulation and the displacement map reconstructed by the DHM, both highlighting the probe points where the displacement was recorded; and f) Out-of-plane displacement measured by the DHM in the two selected probe points.

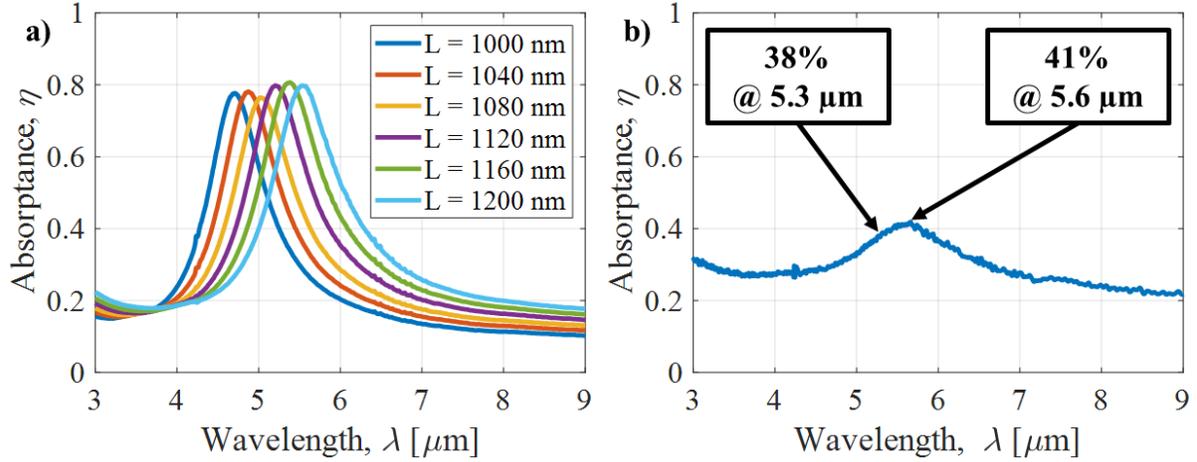

Figure 5: a) IR absorptance spectra obtained via FTIR measurements of the designed metasurface fabricated on the test structures with the same material stack as the nanoplate and its tuning obtained by sweeping the length L of the cross, while keeping both its periodicity P and width W equal to 1.5 *μm* and 300 nm, respectively; and b) IR absorptance spectrum obtained via FTIR measurement of the designed metasurface on the device.

A further reduction of the theoretically achievable NEP can be realized by reducing the noise-induced frequency fluctuations through both increasing the resonator quality factor and improving its power handling, together with boosting the IR responsivity through the enhancement of both the thermal resistance and the absorptance of the plasmonic metasurface. On the other hand, an appropriate scaling of the device footprint and material stack thicknesses has to be performed to simultaneously reduce the thermal capacitance and keep the value of the thermal time constant still reasonably low.

## Acknowledgement

The authors thank the staff at both the Northeastern University George J. Kostas Nanoscale Technology and Manufacturing Research Center and Harvard University Center for Nanoscale System for their help. We would also like to thank the DARPA OpTIm program for funding this research project.